\documentclass{article}

\usepackage[preprint]{neurips_2021}

\usepackage[utf8]{inputenc} 
\usepackage[T1]{fontenc}    
\usepackage{hyperref}       
\usepackage{url}            
\usepackage{booktabs}       
\usepackage{amsfonts}       
\usepackage{nicefrac}       
\usepackage{microtype}      
\usepackage{xcolor}         
\usepackage{hyperref}
\usepackage{url}

\usepackage{amsfonts}
\usepackage{amsmath}
\usepackage{amssymb}

\usepackage{graphicx}

\usepackage[nomargin,inline,final]{fixme}
\usepackage{multirow}
\usepackage{wrapfig}
\usepackage{subfig}

\title{Microdosing: Knowledge Distillation for GAN based Compression}

\author{%
	  Leonhard Helminger \\
	  Department of Computer Science \\
	  ETH Zurich, Switzerland\\
	  \And
	  Roberto Azevedo\\
	  DisneyResearch|Studios \\
	  \And
	  Abdelaziz Djelouah\\
	  DisneyResearch|Studios \\
	  \AND
	   Markus Gross \\
	   Department of Computer Science \\
	   ETH Zurich, Switzerland\\
	   \And
	   Christopher Schroers \\
	   DisneyResearch|Studios \\
}
\begin{document}

\maketitle
\begin{abstract}
	
	Recently, significant progress has been made in learned image and video 
	compression. 
	In particular, the usage of Generative Adversarial Networks has led to
	impressive results in the low bit rate regime.
	%
	However, the model size remains an important issue in current 
	state-of-the-art proposals, and
	existing solutions require significant computation effort on the decoding side.
	This limits their usage in realistic scenarios and the extension to video
	compression.
	In this paper, we demonstrate how to leverage knowledge distillation to obtain
	equally capable image decoders at a fraction of the original number of
	parameters.
	We investigate several aspects of our solution including sequence
	specialization with side information for image coding. 
	Finally, we also show how to transfer the obtained benefits into the setting of 
	video compression. 
	Altogether, our proposal allows to reduce a decoder model size by a factor of
	20 and to achieve 50\% reduction in decoding time.
\end{abstract}

\section{Introduction}
\label{sec:intro}
Initially, projections for 2022 have estimated that video content would reach 82\% of internet traffic.
While this would already make video the most prevalent form of media in terms
of bandwidth by far, the current pandemic situation pushed video traffic even
further than these expectations due to the need for social distancing and home officing \citep{url:corona}.
As a result, compression techniques are more challenged than ever to handle
visual data efficiently, and improvements can impact the daily lives of 
millions of people. 

In contrast to the hand-crafted individual components of traditional codecs,
learned image compression schemes aim to learn an optimal non-linear transform
from data, ideally in an end-to-end fashion.
At a high level, most of the methods can be understood as a sort of
\emph{generative model} that tries to reconstruct the input instance
from a quantized latent representation, coupled with a prior that is used to
compress these latents losslessly \citep{balle2017endtoend}. 
Although providing good perceptual quality in the high bitrate target setting,
it is the low bitrate setting in which neural image compression has shown most
of its strength.
In particular, GAN~(Generative Adversarial Networks)-based architectures for
image compression~\citep{agustsson2019extreme,mentzer2020hific} are able to
produce impressive results by generating an appropriate hallucination of detail in the
output image.
As a drawback, GAN-based compression frameworks usually have large decoder
models that are many times trained on private datasets.
Therefore, retraining these models to their original performance is not
generally possible, and even if the data was present, it would not be
straightforward and time consuming.
In addition, the memory requirements and inference time make them less
practical, especially in the context of video coding and mobile devices.

This paper proposes a knowledge distillation~(KD)~\citep{beyer2021knowledge}
approach that allows retaining good perceptual image quality 
while reducing the size of the decoder.
The goal of KD is to transfer the learned knowledge of a 
\emph{teacher network} on a smaller \emph{student network} that remains
competitive to the teacher network performance.
By requiring less memory and computational power than the initial teacher
network, the student network could, for instance, run on less powerful devices
such as mobile phones or dedicated devices.
Being able to compress the generator network or decoder in the auto-encoder
setting is not only interesting in terms of memory requirements but also in terms of computational efficiency.
This is especially important for image and video compression, where the
majority of the computation should preferably be on the sender~(encoder) side,
while the decoding should be simple. Especially in the context of video streaming, an asset will typically be encoded once while it will be distributed and decoded millions of times. Traditional codecs do spend a lot of emphasis and compute on encoding while keeping decoding extreme lightweight. In contrast to that, deep learning based architectures are typically symmetric in terms of encoding and decoding or can even require considerably more compute for decoding when using a GAN~\citep{mentzer2020hific}.

Our proposal is based on:
i)~training a reduced \emph{student decoder} with data generated from the
\emph{big decoder}.
ii)~overfitting the reduced \emph{student decoder} model to a specific image or
set of images; and
iii)~sending the specialized decoder weights alongside the image latents.
To show the viability of our proposal, we incorporate it into state-of-the-art
models for neural image and video compression targeting the low bitrate setting.
First, we replace the \textbf{High-Fidelity
Compression~(\emph{HiFiC})}~\citep{mentzer2020hific} decoder with a much smaller
student decoder.
\emph{HiFiC} is the state-of-the-art in low bitrate neural image
compression~(\textasciitilde 0.15 bpp) that produces extremely competitive results at the cost of a fairly big (\textasciitilde 156M parameters) decoder network.
Our proposed KD approach allows for a much smaller decoder~(\textasciitilde 8M
parameters) and 50\% faster decoding time while still producing visually
similar output images.
Second, we show how to apply our KD strategy in a neural video compression framework based
on latent residuals~\citep{djelouah2019vidcodec}.
In such a scenario, we overfit our reduced \emph{student decoder} to a sequence
so that we can provide a sequence specific decoder.
We show that the additional bits needed for sending the weights amortize over the sequence.

Explicitly, the main contributions of this paper are: i)~ proposing novel
strategies for KD for neural image and video compression;
ii)~investigating KD in the low bitrate setting for GAN-based image
compression;
iii)~investigating KD in the low bitrate GAN-based video compression setting
with latent residuals.


\section{Related Work}
\label{sec:related_work}

\paragraph{Neural image compression}
The first proposed neural image compression
methods~\citep{balle2016end-to-end,toderici2015variable,toderici2017full} showed
improved results over JPEG or JPEG2000, while most recent
approaches~\citep{balle2018variational,mentzer2018conditional,minnen2018full,helminger2020lossy,DBLP:conf/iccv/ChoiEL19}
are now on par or surpassing BPG~\citep{nxp:bpg}.
Recent
works~\citep{DBLP:journals/corr/abs-1908-04187,Patel_2021_WACV,DBLP:journals/corr/abs-1907-08310,DBLP:journals/corr/abs-2106-04427}
also investigate the application of perceptual losses and how they relate to the
human visual perception of decompressed images.

In particular, generative methods have been providing impressive results in the
low bitrate setting~\citep{agustsson2019extreme,mentzer2020hific}.
Generative models for image compression are able to synthesize details that
would be very costly to store, resulting in visually pleasing results at
bitrates in which previous methods show strong
artifacts~\citep{agustsson2019extreme}.
However, to synthesize those details, a powerful and potentially big decoder is
necessary, which contradicts the general requirements on compression
technologies, where the sender~(encoder) should have the burden of computing a
good compression, such that the receiver~(decoder) can be as simple as
possible.
Our approach can be applied to any of the neural image compression methods
above, whenever the decoder is too big and faster inference times are required.
This is especially crucial in GAN-based neural compression methods.
As a use case, we show an application of our proposal using
\emph{HiFiC}~\citep{mentzer2020hific}, which is the state-of-the-art GAN model for
low bit rate image compression.

\paragraph{Neural video compression}
As an extension to the neural image compression methods, neural video
compression approaches aim to leverage redundancy in both spatial and temporal 
information.
\citep{lu2018dvc} replace blocks in traditional video codecs with neural
networks.
Learning-based optical flow estimation is used to obtain the motion information
and to reconstruct the current frames.
Then they employ two auto-encoder networks to compress the
corresponding motion and residual information.
\cite{veerabadran2020adversarial} show that minimizing an
auxiliary adversarial distortion objective for neural video compression in the low
bitrate setting creates distortions that better correlate with human perception.
\cite{djelouah2019vidcodec} propose compressing and sending
residuals in the latent space instead of residuals in the pixel space, which
allows the reuse of the same image compression network for both keyframes and
intermediate frames.
In this paper, we use such an approach as a starting point to show the
viability of KD in neural video compression.
Instead of explicitly computing residuals or differences, \cite{ladune2021conditional} use feature space concatenation and train a
decoder that operates on this joint information. 
\cite{van2021overfitting} present an extreme approach that
fine tunes the full model to a single video, and sends model updates (quantized
and compressed using a parameter-space prior) along with the latent
representation.
Such an approach is in line with our idea of sequence-specific information to
be sent to the student decoder, and our work can be seen as complementary
to~\citep{van2021overfitting}, which does not include KD.

\paragraph{Knowledge distillation}
KD has been primarily used on vision tasks like object classification or
segmentation.
KD for generative models, however, is not well studied yet.
In~\citep{chang2020tinygan}, the authors leveraged a teacher-student architecture
to reduce the size of the BigGAN \citep{brock2018large} architecture while still
being competitive on Inception and FID scores.
To the best of our knowledge, our work is the first to propose knowledge
distillation to learn a smaller decoder for neural image and video compression frameworks.

\section{Knowledge Distillation for Compression}
\label{sec:methods}

\begin{figure}
\centering
\includegraphics[width=0.80\textwidth]{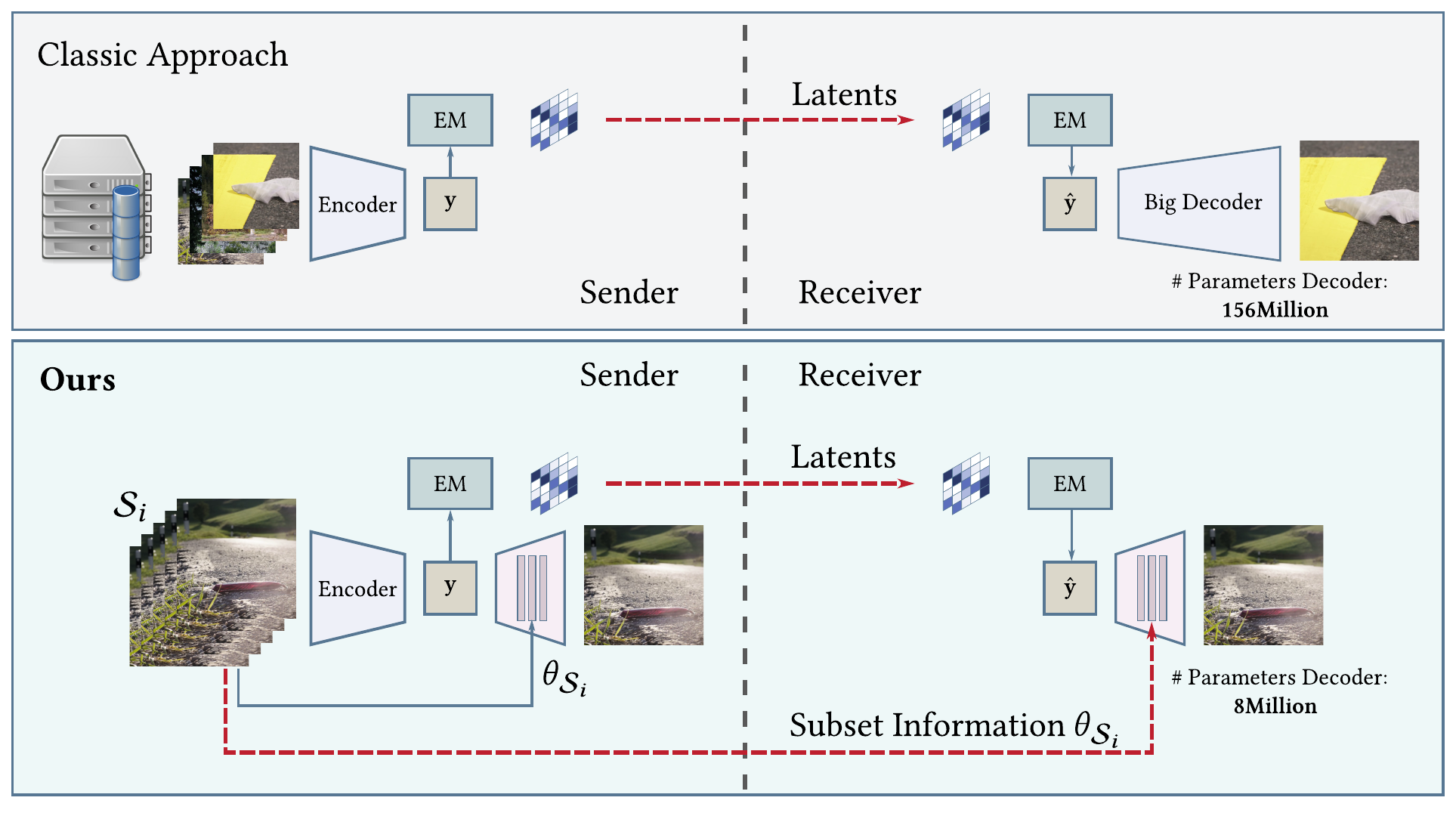}
\caption{Overview of our proposal: we replace the Big Decoder of GAN
	compression models (top) with a smaller decoder plus a
        content-specific additional information~(bottom).}
\label{fig:overview}
\end{figure}

Figure~\ref{fig:overview} illustrates an overview of our proposal.
In the classic neural compression approach, the encoder-decoder pair is
trained on a big dataset, to get an overall good performance on a variety of
different content.
Once the auto-encoder is fully trained, the decoder gets deployed and sent to
the receiver.  The potentially big decoder then allows to decode any type of
content.
In our approach, we enable the sender to partition the data into subsets
$\mathcal{S}_i$, and learn a content-specific decoder with corresponding
information $\theta_{\mathcal{S}_i}$ for each subset.
This specialization allows us to train a model with less parameters, smaller
memory footprint, and less computations.
Once the decoder is fully trained, and the sender's reconstruction quality
requirement of the subset is fulfilled, the content-specific information is
stored alongside the subset.
If the receiver wants to decode an image $\mathbf{x} \in \mathcal{S}_i$, the
subset specific information $\theta_{\mathcal{S}_i}$ has to be sent once per 
subset.
Next, we detail how to apply our distillation process to image compression
with GANs and extend this to video compression using latent space residuals.

%
%

\subsection{Knowledge Distillation for Image Compression with GANs}

\paragraph{High-Fidelity Generative Image Compression~(HiFiC)}

\begin{figure}[tb]
\centering
\includegraphics[width=0.9\textwidth]{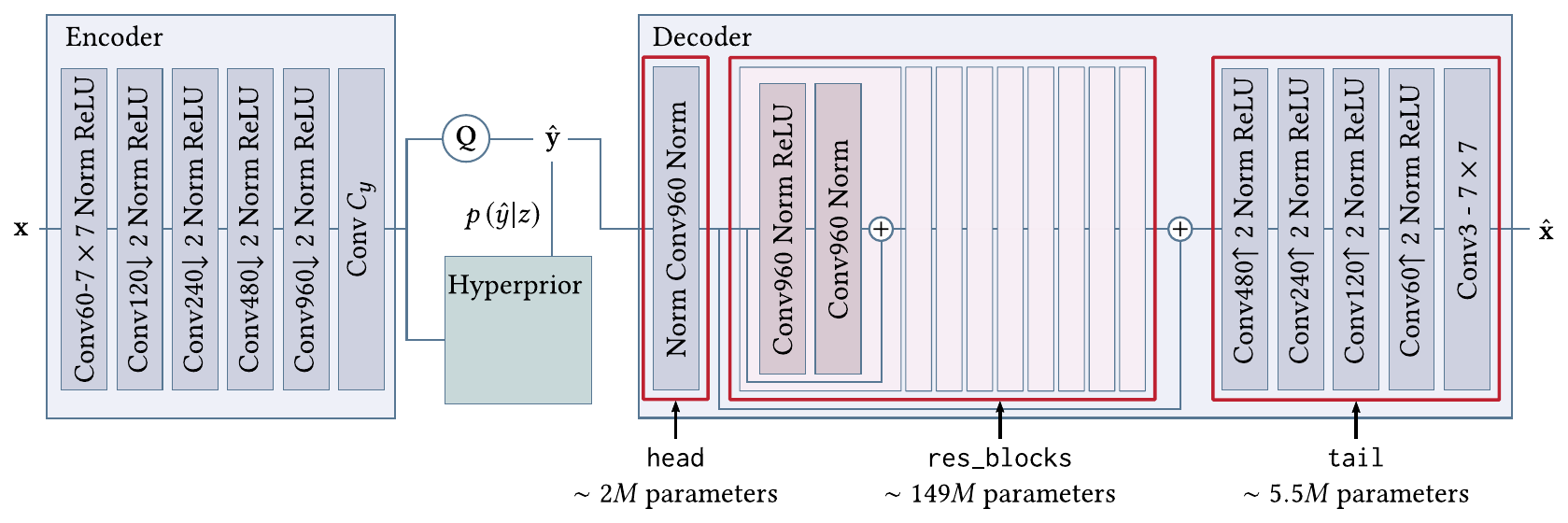}
\caption{Architecture of High-Fidelity Generative Image Compression (\emph{HiFiC})}
\label{fig:hific}
\end{figure}

\begin{figure}[b]
\centering
\subfloat[$\text{\emph{HiFiC}}^{Lo}$]{
  \includegraphics[width=0.23\textwidth]{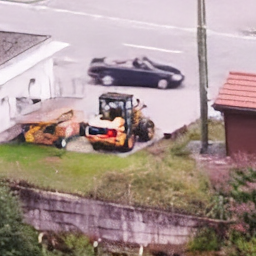}
}
\subfloat[$\text{\emph{HiFiC}}^{Lo}$ (w/o RB)]{
	\includegraphics[width=0.23\textwidth]{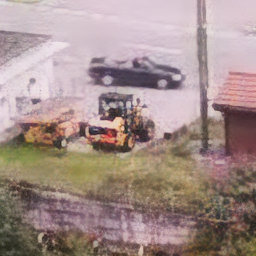}
}
\subfloat[$\text{\emph{HiFiC}}^{Hi}$]{
	\includegraphics[width=0.23\textwidth]{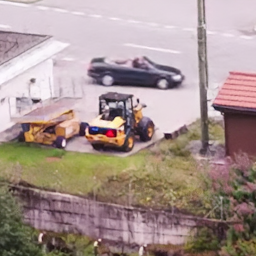}
}
\subfloat[$\text{\emph{HiFiC}}^{Hi}$ (w/o RB)]{
  \includegraphics[width=0.23\textwidth]{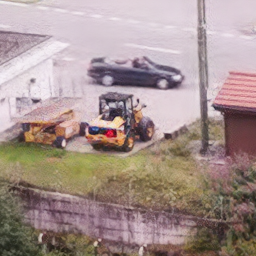}
}
\caption{Even without residual blocks (RB), the \emph{HiFiC} model outputs a coarse version of the image.
The images show the output of \emph{HiFiC} models \textbf{with} (a), (c) and \textbf{without} (b), (d) residual blocks. }
\label{fig:hific_resblocks}
\end{figure}

Figure~\ref{fig:hific} shows the HiFiC architecture~\citep{mentzer2020hific}
\footnote{The authors of \citep{mentzer2020hific} trained their architecture for
different target bit rates and provide these models as: $\text{\emph{HiFiC}}^{Hi}$,
$\text{\emph{HiFiC}}^{Mi}$ and $\text{\emph{HiFiC}}^{Lo}$.}.
Its decoder can be divided into three sub-nets:
\emph{head}~(\textasciitilde 2M parameters),
\emph{res\_blocks}~(\textasciitilde 149M parameters), and
\emph{tail}~(\textasciitilde 5.5M parameters).
Through experimentation, and as shown in
Figure~\ref{fig:hific_resblocks}, it is easy to conclude that the coarse
information of the image is saved in the latent space, and the hallucination of
the texture is generated by the residual network (\emph{res\_blocks}) of the
decoder.
In particular, the discrepancy in size of the the \emph{res\_blocks} is due to
the fact that the model was trained on a big~(private) dataset, thus such a big
size is needed to capture all the textures seen during training.
However, if we know in advance which images should be compressed (e.g. frames of a
video with similar features), we can overfit to this data and sent only the
necessary weights to properly decode these images.
That is exactly what we propose with the Distilled-HiFiC architecture bellow.

\begin{figure}[tb]
	\centering
	\subfloat[Teacher-Student Architecture\label{fig:teacher_student}]{
		\includegraphics[width=0.45\textwidth]{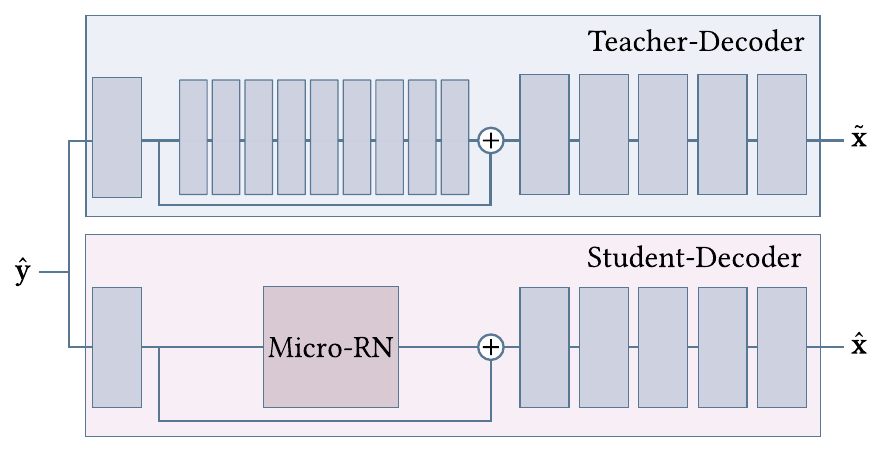}
	}\qquad\qquad
	\subfloat[Micro-RN\label{fig:dablock}]{
		\includegraphics[width=0.4\textwidth]{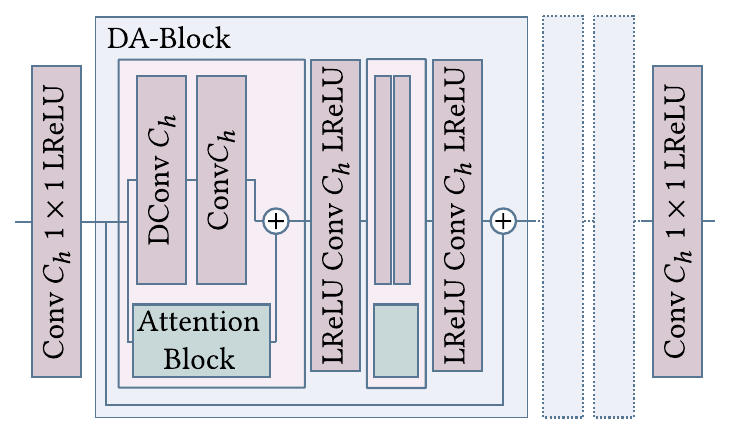}
	}
	\caption{Visualization of the proposed \emph{Micro-RN}. 
	If not stated explicitly, a $3\times3$ convolution is used.
	DConv denotes a depthwise convolution and $C_h$ is the number of hidden channels.}
\end{figure}

\paragraph{Distilled-HiFiC} 
Our proposed Distilled-HiFiC (See Figure~\ref{fig:teacher_student}) reduces the
size of the decoder by training a smaller sub-network, named
\emph{Micro-Residual-Network~(Micro-RN)} that mimics the behavior of the
residual network~(\emph{res\_blocks}) for a specific subset, and therefore
\textit{microdosing} the capability of hallucinations.
Micro-RN is based on the \emph{degradation-aware}~(DA) blocks introduced in
\citep{DBLP:journals/corr/abs-2104-00416}.
While \citep{DBLP:journals/corr/abs-2104-00416} utilizes a kernel prediction network to steer the weights according
to a degradation vector, we learn a set of weights $\theta_{\mathcal{S}_i}$ per
subset $\mathcal{S}_i$.
Micro-RN is defined by two parameters: $C_h$, the number of hidden channels, and
$B$, the number of DA Blocks.
Similar as in \citep{chang2020tinygan}, we train Micro-RN with the
teacher-student architecture (Figure \ref{fig:teacher_student}), with the
difference that our decoder borrows pre-trained layers (i.e., \emph{head} and
\emph{tail}) from the teacher-decoder.
Let $\mathbf{x} \in \mathcal{S}_i$ be an image of subset $\mathcal{S}_i$ and
$\tilde{\mathbf{x}}$ be the image compressed by the teacher network.
We optimize the following loss:
\begin{equation}
	\mathcal{L}\left(\mathbf{x}; \theta_\mathcal{S} \right) = k_M \text{MSE}\left(\tilde{\mathbf{x}}, \hat{\mathbf{x}}\right) + k_p d_p\left(\hat{\mathbf{x}}, \mathbf{x}\right),
	\label{eq:kdloss}
\end{equation}
where $\hat{\mathbf{x}}$ is the output of the student network, MSE and $d_p$
are the distortion losses, and $k_M$ and $k_p$ are their corresponding weights.
Similar to \citep{mentzer2020hific} as the perceptual loss
$d_p=\text{LPIPS}$~\citep{zhang2018unreasonable} is used. 
Hence, the loss forces the student-decoder to generate images that look
similar to the teacher's and further reduce the perceptual loss to the ground
truth image.
Note, that we freeze both the encoder and the entropy model which in this case is modeled using a hyperprior.
Hence, $\theta_\mathcal{S}$ only contains the weights of the Micro-RN.
This allows us to leverage the powerful encoder and hyperprior of $\text{\emph{HiFiC}}$ as well as
the models knowledge about the private training data set.
 %
%

\subsection{Knowledge Distillation for Video Compression with Latent Space Residuals}

\begin{figure}[t]
\centering
\includegraphics[width=0.95\textwidth]{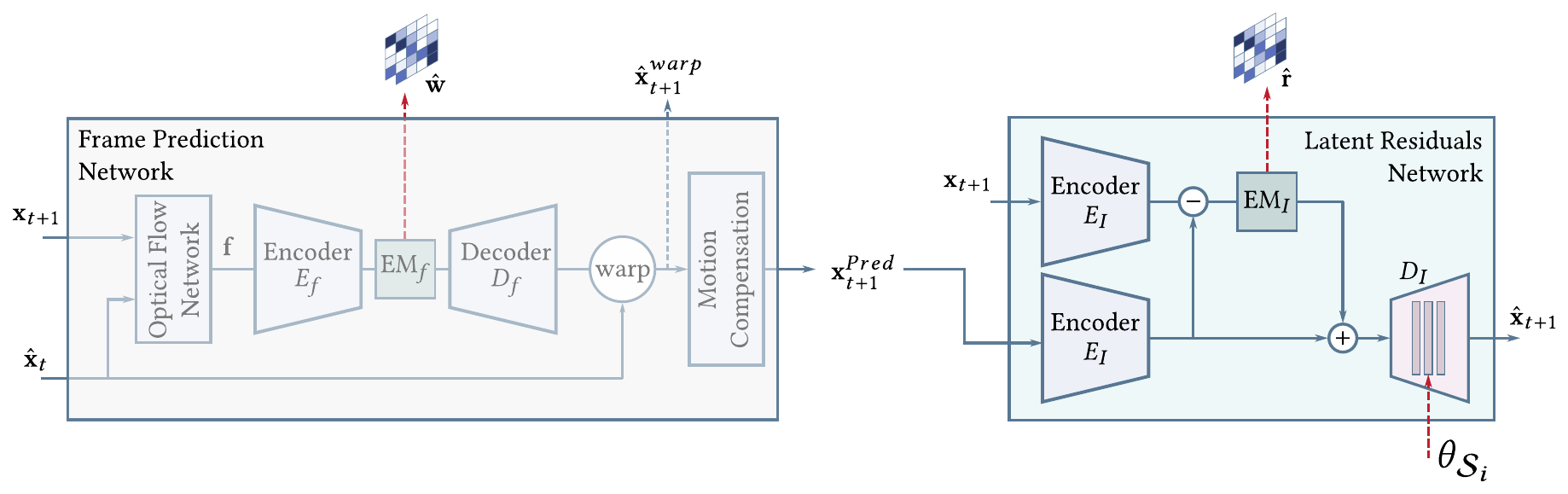}
\caption{Overview of knowledge distillation with latent residuals for video
         compression: Frame Prediction Network (FPN) and Latent Residual Network (LRN)}
\label{fig:video_compression_overview}
\vspace{-1em}
\end{figure}

%

To show the application of KD in neural video compression scenarios, we use a
similar approach as proposed by \cite{DBLP:conf/iccv/DjelouahCSS19}.%
\footnote{The main differences of our implementation compared to 
\citep{DBLP:conf/iccv/DjelouahCSS19} are:
i)~we only use P-Frames, while~\citep{DBLP:conf/iccv/DjelouahCSS19} also uses
B-Frames;
ii)\citep{DBLP:conf/iccv/DjelouahCSS19} retrains the image
encoder/decoder from scratch, while we reuse the pre-trained HiFiC
encoder/decoder during training time, and replace the HiFiC decoder with our
Distilled-HiFiC during inference time;
iii)~we start with a pre-trained optical flow and allow the flow to be fine
tuned during the training of the latent residuals network.}~(see
Figure~\ref{fig:video_compression_overview}).
Such network is composed of two parts: a \emph{Frame Prediction
Network~(FPN)} and a \emph{Latent Residual Network~(LRN)}.
Given a sequence of frames~(\emph{group of pictures}, or GOP) 
to be encoded
$\mathbf{x}_0,...,\mathbf{x}_{GOP}$, where $\mathbf{x}_0$ is a keyframe~(I-frame) and $\mathbf{x}_1,...,\mathbf{x}_{GOP}$ are predicted frames~(P-frames), the compression of the sequence works as follows:

First, the I-frame~($\mathbf{x}_0$) is compressed using a neural
image-only compression network
to generate the encoded latent~$\mathbf{y}_0$. $\hat{\mathbf{x}}_0$ denotes
the reconstructed frame from the quantized latent $\hat{\mathbf{y}}_0$.
Then, for each P-frame, $\mathbf{x}_{t+1}, 1 \leq t+1 \leq GOP$, we:
(1) generate a temporal prediction, $\mathbf{x}^{Pred}_{t+1}$, of
$\mathbf{x}_{t+1}$ from the previous reconstructed frame,
$\hat{\mathbf{x}}_{t}$, using \emph{FPN}.
The \emph{FPN} works by first computing the optical flow $\mathbf{f}_{t+1}$
between $\mathbf{x}_{t+1}$ and $\hat{\mathbf{x}}_{t}$.
(2) use the neural motion compression network to generate the encodings and quantized latents $\hat{\mathbf{w}}_{t+1}$ of $\mathbf{f}_{t+1}$.
(3) warp $\hat{\mathbf{x}}_{t}$ with the decompressed flow $\hat{\mathbf{f}}_{t+1}$, and then motion compensating it to generate the temporal $\mathbf{x}^{Pred}_{t+1}$.\\
To compute the residual between the temporal prediction and the P-Frame, we use the \emph{LRN}:
(4) we encode both, the prediction $\textbf{x}^{Pred}_{t+1}$ and $\textbf{x}_{t+1}$, with $E_I$ (a pretrained image compression encoder) and
(5) compute the latent residual, $\mathbf{r}_{t+1}$, between the latents of
the P-frame against the predicted frame,
$\mathbf{r}_{t+1}=\mathbf{y}_{t+1}-\mathbf{y}^{Pred}_{t+1}$, 
which is then quantized and entropy coded with ~EM$_I$.
The final bitstream of a GOP is then composed of $\{\hat{\mathbf{y}}_0, \hat{\mathbf{w}}_1,
..., \hat{\mathbf{w}}_{GOP},\hat{\mathbf{r}}_1, ...\hat{\mathbf{r}}_{GOP}\}$, i.e., latent of
the I-frame and the compressed flow fields and latent residuals for each of the
P-frames~(all quantized and entropy encoded)\footnote{Details on the different network modules and training procedures are provided in the Appendix}.

In the low bitrate setting, \emph{HiFiC} would seem like a suitable choice for the neural image compression
architecture that could be used together with the above latent space residual
framework.
As previously mentioned, however, the size of the \emph{HiFiC} decoder is a limiting
factor.
Also, inference time is even more important in the video setting that should be
able to keep a decoding frame rate of \textasciitilde 30 frames per
seconds~(fps).
Thus, we propose to use our Distilled-HiFiC in the latent residual framework
above.

During encoding, we overfit our Distilled-HiFiC to a specific sequence so that
we only need to send the $\theta_{\mathcal{S}}$ once for all the frames of that
sequence.
Our decoding process starts then by receiving and loading sequence-specific
Micro-RN weights on the Distilled-HiFiC decoder, which is then fixed during the
decode of the whole sequence.
As detailed next, with such an small overhead, we can reduce the
decoding time by half while keeping similar visual quality to the latent residual
framework using the original \emph{HiFiC} decoder.

\section{Experiments}
\label{sec:experiments}
In this section we first describe the experiment that leads
to the minimal architecture that is powerful to mimic \emph{HiFiC}'s residual network on a variety of subsets.
In the second part of this section, we show that our student decoder can be
used for video compression based on latent space residuals and compare the results with
OpenDVC, a recent neural video compression method.
 
\paragraph{Dataset} The models were trained on two types of subsets: 
$\mathcal{S}_{uvg}$ a subset of UVG \citep{10.1145/3339825.3394937} and 
$\mathcal{S}_{i}$ a subset of sequence $i \in \{1, \dots, 7\}$. 
We created the subsets by taking every 10th frame of each clip.
This allows Micro-RN to learn the image features present in the whole clip.
$\mathcal{S}_{uvg}$ consists in total of 390 frames with resolution
$1920\times 1080$ and 
$\mathcal{S}_{i}$ of up to 60 frames. 

We evaluated the proposed model for qualitative comparisons on an in-house created movie. The subset $S_i$ corresponds to a specific clip, where $S_{lucid}$ denotes the union of four clips.

\subsection{Micro-Residual-Network}

\paragraph{Training} We trained our Micro-RN by minimizing
Equation~\ref{eq:kdloss} on random crops of size $256\times256$.
We used the hyper parameters $k_p$ and $k_M$ proposed in
\citep{mentzer2020hific}.
We further used the Adam optimizer \citep{DBLP:journals/corr/KingmaB14} with
a learning rate of $10^{-4}$ and batch size 4. Training a model takes approximately 10
hours on a NVIDIA Titan Xp.

\paragraph{Ablation Study}
To find the minimum number of weights for Micro-RN, we trained our student
decoder with various number of hidden channels $C_h = \{64, 128, 256\}$ and DA
blocks $B=\{1, 2, 3, 4\}$.
We are interested in the smallest possible architecture that is still capable
to mimic $HiFiC$'s residual network.
$HiFiC^{Lo}$ is selected as the teacher network, since it is the model that
has the most difficult task of hallucinating details.
Table~\ref{tab:params} shows the number of parameters per configuration.

\begin{wraptable}{r}{0.4\textwidth}
	\begin{center}
		\vspace{-1em}
		\resizebox{0.27\columnwidth}{!}{%
			\begin{tabular}{@{}ccccc@{}}
				\toprule
				& & \multicolumn{3}{c}{$C_h$} \\
				& & 64 & 128 & 256 \\ 
				\midrule 
				\multirow{4}{*}{\rotatebox[origin=c]{0}{$B$}} & 1 & 211k & \textbf{594k} & 1.88M \\ 
				& 2 & 294k & 925k & 3.19M \\ 
				& 3 & 378k & 1.25M & 4.51M \\ 
				& 4 & 461k & 1.59M & 5.82M \\ 
				\bottomrule
			\end{tabular}
		}
		\caption{Number of parameters per tested configuration.}
		\label{tab:params}
	\end{center}
	\vspace{-2em}
\end{wraptable}

\begin{figure}
	\centering
	\subfloat[Beauty\label{fig:model_comparison_beauty}]{
		\includegraphics[width=0.26\textwidth]{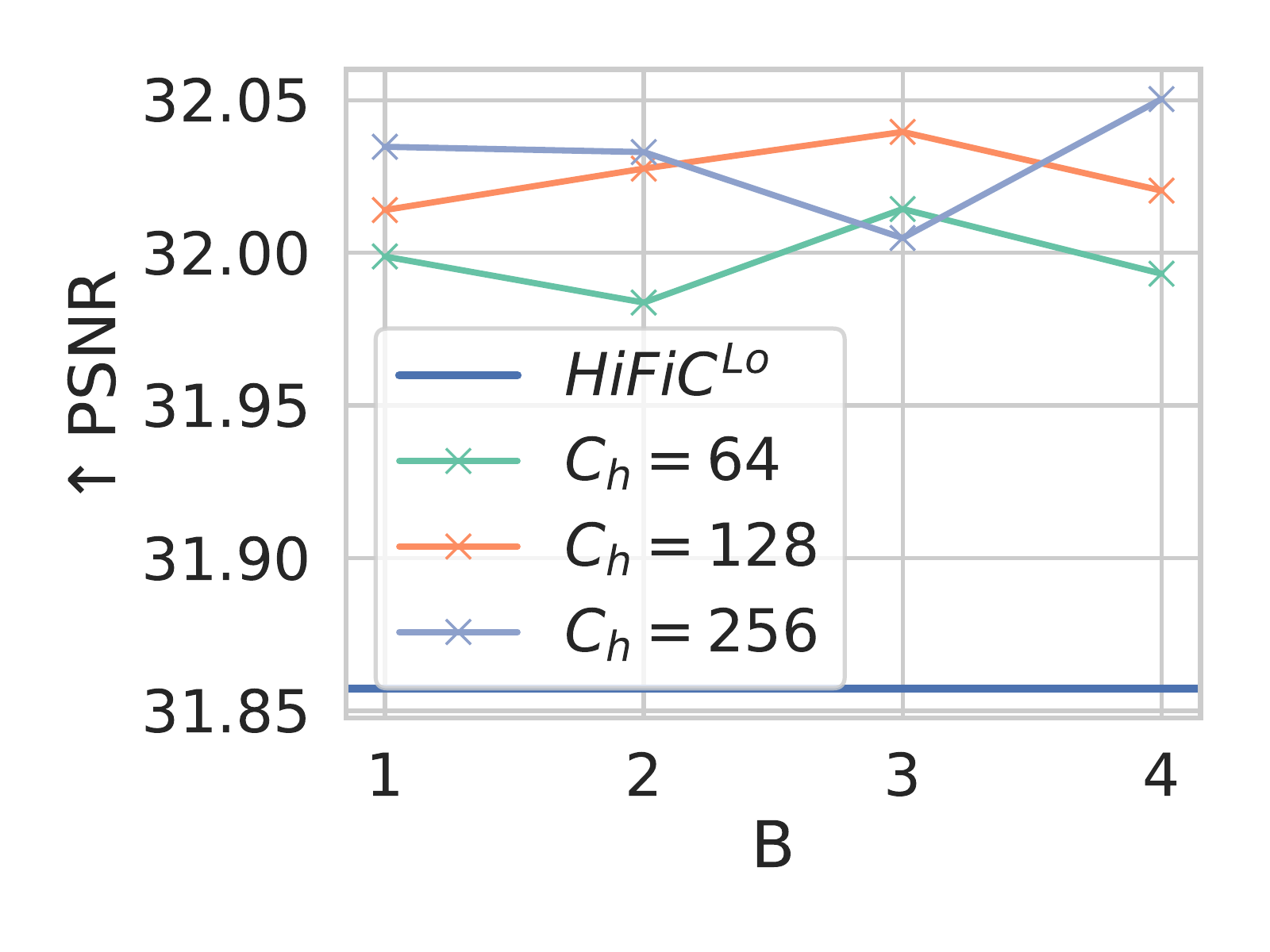}\hspace{-1em}
		\includegraphics[width=0.26\textwidth]{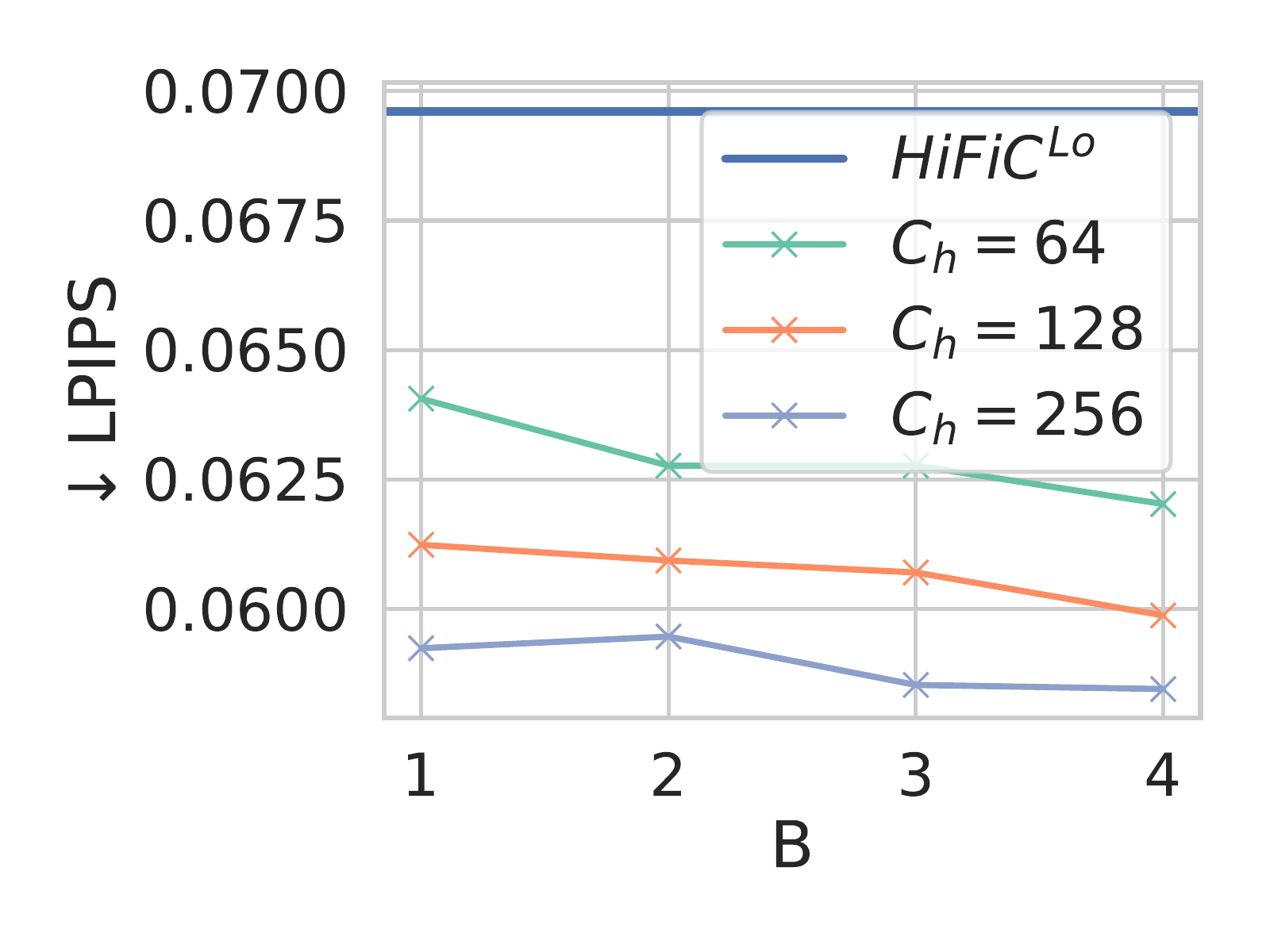}
	}~\subfloat[\label{fig:model_comparison_yacht}YachtRide]{
		\includegraphics[width=0.26\textwidth]{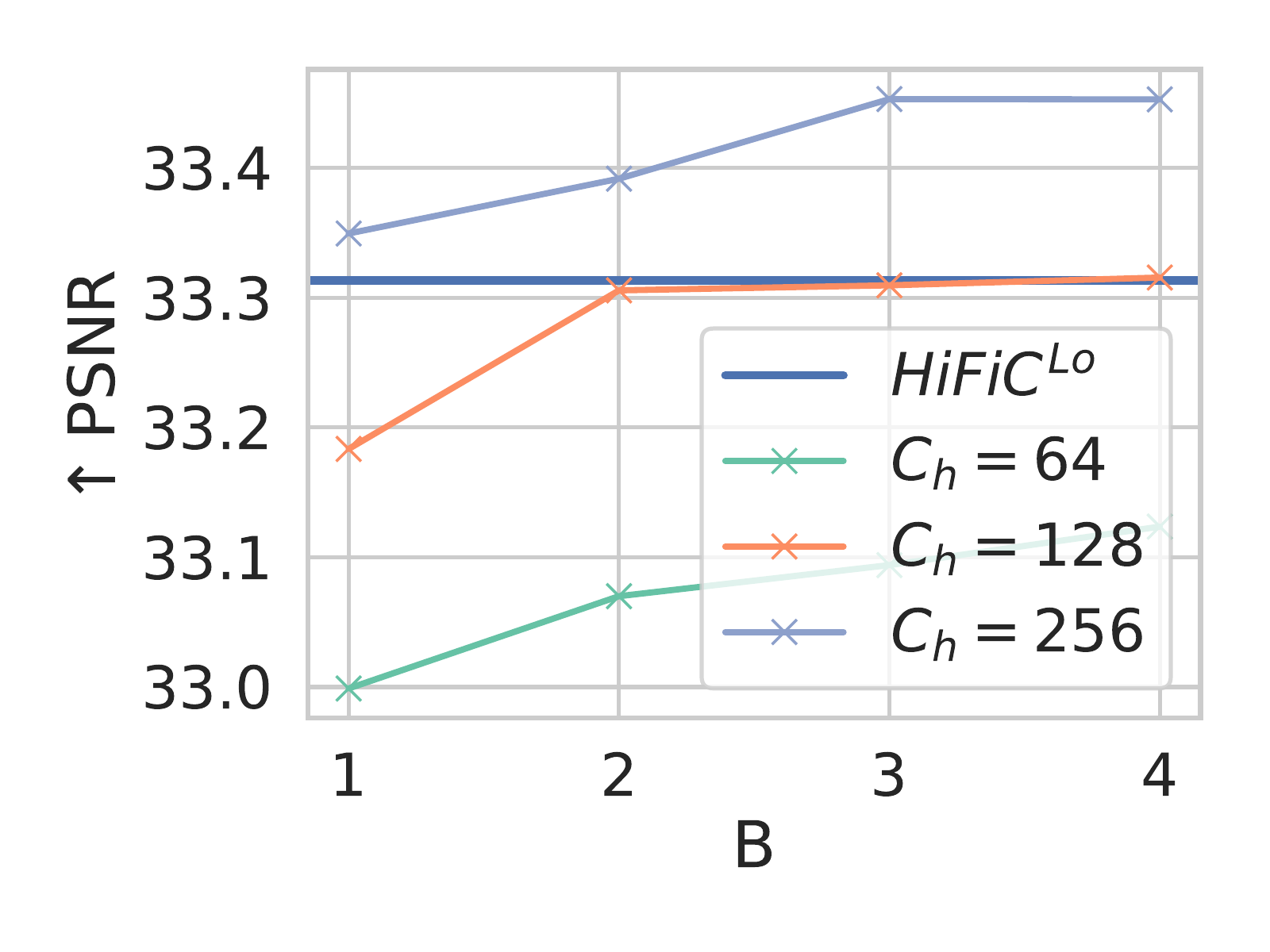}\hspace{-1em}
		\includegraphics[width=0.26\textwidth]{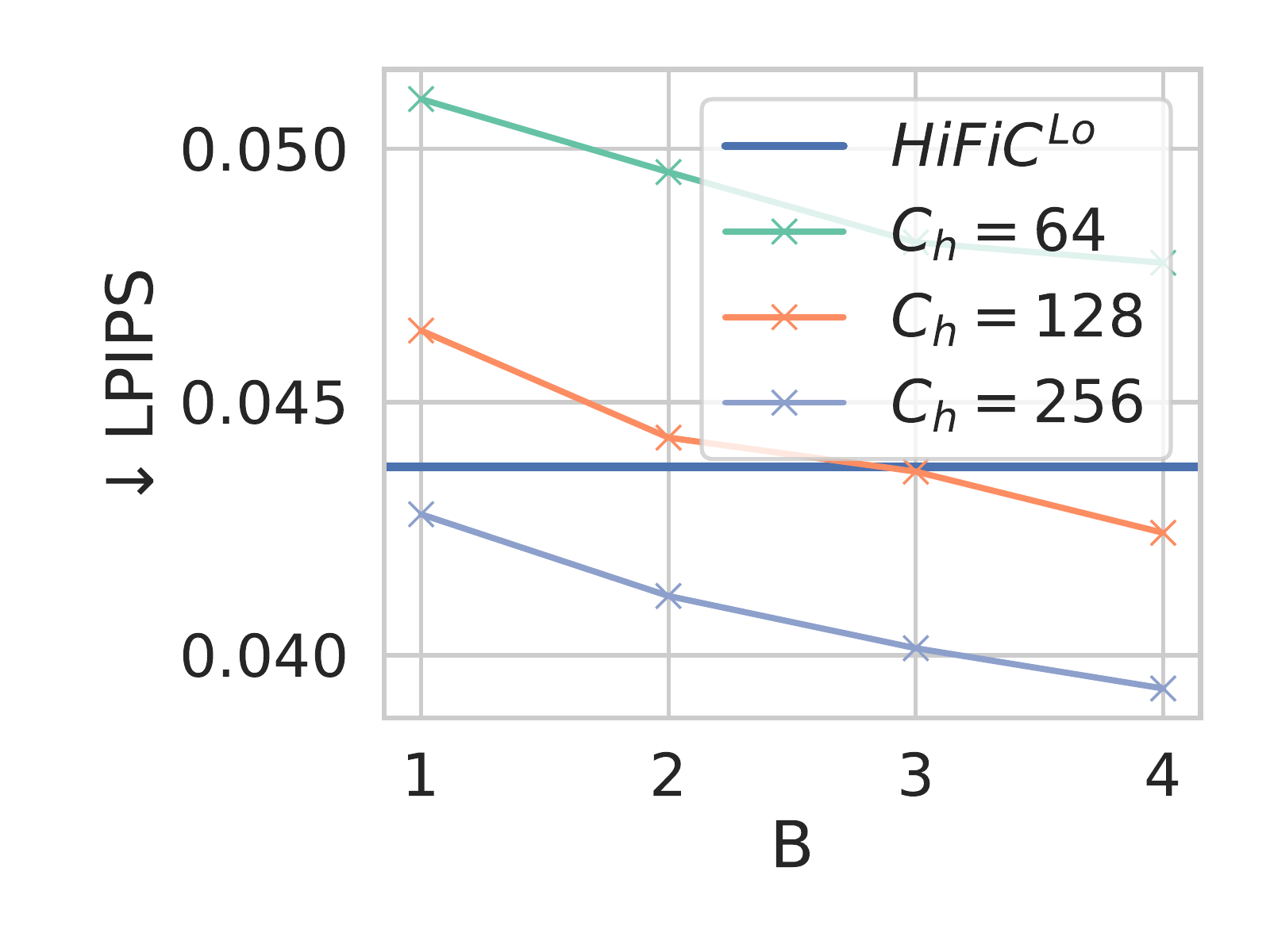}
	}
	\caption{Ablation Study: We show that increasing the number of hidden channels $C_h$ or number of DA blocks $B$ allows to achieve a similar performance as $\text{\emph{HiFiC}}^{Lo}$.}
	\label{fig:model_comparison}
\end{figure}

\paragraph{Architecture}
To find the minimum number of weights for Micro-RN, we trained our student
decoder with various number of hidden channels $C_h = \{64, 128, 256\}$ and DA
blocks $B=\{1, 2, 3, 4\}$.
We are interested in the smallest possible architecture that is still capable
to mimic $\text{\emph{HiFiC}}$'s residual network.
$\text{\emph{HiFiC}}^{Lo}$ is selected as the teacher network, since it is the model that
has the most difficult task of hallucinating details.
Table~\ref{tab:params} shows the number of parameters per configuration. 
The experiments show that increasing either $C_h$ or $B$
results in a higher reconstruction quality, and better mimicking the $\text{\emph{HiFiC}}$
decoder. 
Another interesting aspect is that the required number of hidden channels $C_h$ 
and number of DA Blocks $B$
depend on the complexity and the details of a subset.
For less complex sequences e.g. \emph{Beauty}, the smallest tested configuration
with $C_h=64$ and $B=1$ is sufficient to outperform \emph{HiFiC}$^{Lo}$ in terms
of PSNR and LPIPS (see green curve in Figure \ref{fig:model_comparison_beauty}).
Since the sequence consists primarily of black noise background with less features
that need to be reconstructed, the Micro-RN does not need to save much details about the texture.

For sequences with many details and a high variation of features e.g. \emph{YachtRide},
more parameters, i.e., more hidden channels and DA blocks are necessary.
This can be seen in Figure \ref{fig:model_comparison_yacht}. 
The green curve ($C_h=64$) is always below the baseline of $\text{\emph{HiFiC}}^{Lo}$.
By increasing the number of channels to $C_h=128$ (orange curve) and $B=2$ our approach produces similar numbers to the baseline.

%
%

\paragraph{Architecture} Based on these results and for simplicity, we chose to set the number of hidden channels to
$C_h = 128$ and $B = 1$ for all other experiments.
Increasing either $B$ or $C_h$ would lead to a double or almost quadratic
increase of parameters for higher $B$, respectively.

\begin{wraptable}{r}{0.4\textwidth}  
\begin{center}
	\vspace{-1.5em}
	\resizebox{0.3\columnwidth}{!}{%
		\begin{tabular}{@{}lcc@{}}
			\toprule 
			& Decoding [s] \\ 
			\toprule 
			Lower Bound &  0.20 \\
			\midrule
			HiFiC~(pruned) & 0.44  \\ 
			HiFiC~(built) & 0.45  \\ 
			\textbf{Ours} & \textbf{0.20} \\ 
			\bottomrule
		\end{tabular}
	}
\end{center}
\caption{Timings for decoding.}
\label{tab:timings}
\end{wraptable}

\paragraph{Comparison to HiFiC}
We compare Distilled-HiFiC to the original $\text{\emph{HiFiC}}$ decoder on: number of
parameters, decoding time, visual comparisons, and distortion metrics.
Replacing $\text{\emph{HiFiC}}$'s residual network with our proposed
Micro-RN reduces the number of parameters of the decoder from
156M to 8M parameters. 
The proposed Micro-RN itself consists only of 600K parameters.
This does not only reduce the memory footprint of the model, but it would also
allow to decode images of higher resolution in a single pass.
We also conducted a benchmark on the timings for decoding,
for various implementations of $\text{\emph{HiFiC}}$. 
We distinguish between the following models: $\text{\emph{HiFiC}}$, $\text{\emph{HiFiC}}$-(built), $\text{\emph{HiFiC}}$ (w/o ResBlocks) and ours. 
$\text{\emph{HiFiC}}$ is the original metagraph provided by \cite{mentzer2020hific} and
$\text{\emph{HiFiC}}$-(built) is the graph we built ourself during training.
In addition we compare against an incomplete decoder
which serves as a \emph{Lower Bound}.
It has the same architecture as \emph{HiFiC-(built)} but without computing the output of the residual blocks.

Table \ref{tab:timings} shows that decoding with our architecture is twice as
fast as the original one.
It further shows that our model is as fast the lower bound i.e.\ the added
complexity by the Micro-RN is negligible in terms of decoding time.
Since the encoder is the same for every setup and encoding time (\textasciitilde0.23s) is similar for each model.

\begin{figure}
	\centering
	\subfloat[Ground truth]{
		\includegraphics[width=0.22\textwidth]{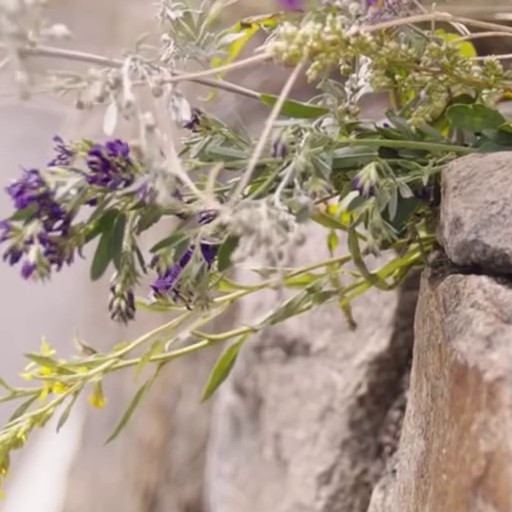}
	}
	\subfloat[Ours $\mathcal{S}_{lucid}$]{
		\includegraphics[width=0.22\textwidth]{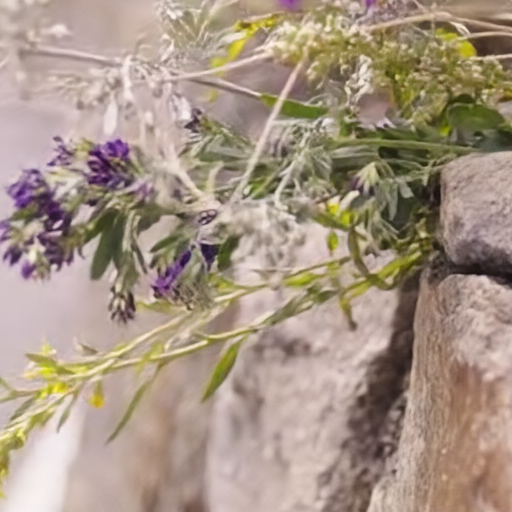}
	}
	\subfloat[Ours $\mathcal{S}_{i}$]{
		\includegraphics[width=0.22\textwidth]{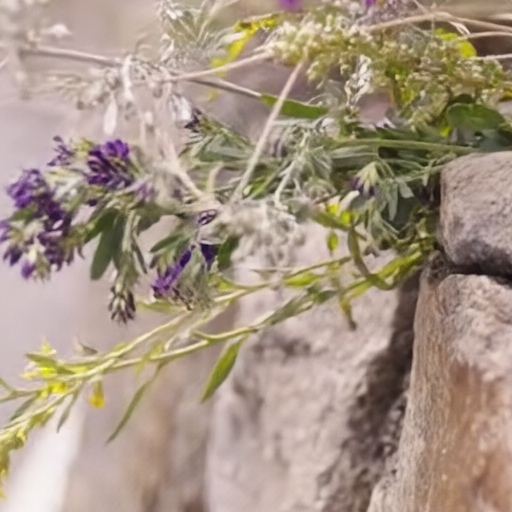}
	}
	\subfloat[$\text{\emph{HiFiC}}^{Lo}$]{
		\includegraphics[width=0.22\textwidth]{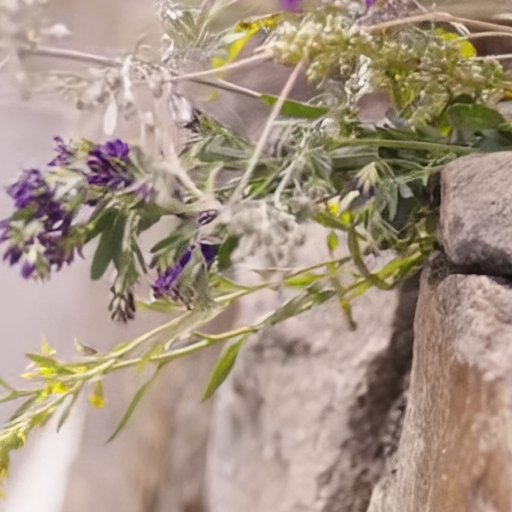}
	}
	\caption{Image Compression: Visual comparison between \emph{HiFiC}$^{Lo}$ and our reduced Decoder.}
	\label{fig:comparison_microrn}
\end{figure}

A visual comparison between the ground truth, a distilled decoder on both 
$\mathcal{S}_{uvg}$ and $\mathcal{S}_{i}$, and $\text{\emph{HiFiC}}^{Lo}$ is provided in
Figure \ref{fig:comparison_microrn}.
Micro-RN is capable to mimic $\text{\emph{HiFiC}}$'s residual network for both subsets, $\mathcal{S}_{uvg}$ and $\mathcal{S}_{i}$.
Further, if we compare both models, the model which was trained only on the sequence, i.e. $\mathcal{S}_{i}$, 
learns more details and better adapt to the specific sequence. 
It also seems that Micro-RN removes LPIPS specific patterns and has a slightly smoother reconstruction.
\begin{figure}%
	\centering
	\subfloat{
		\includegraphics[width=0.31\textwidth]{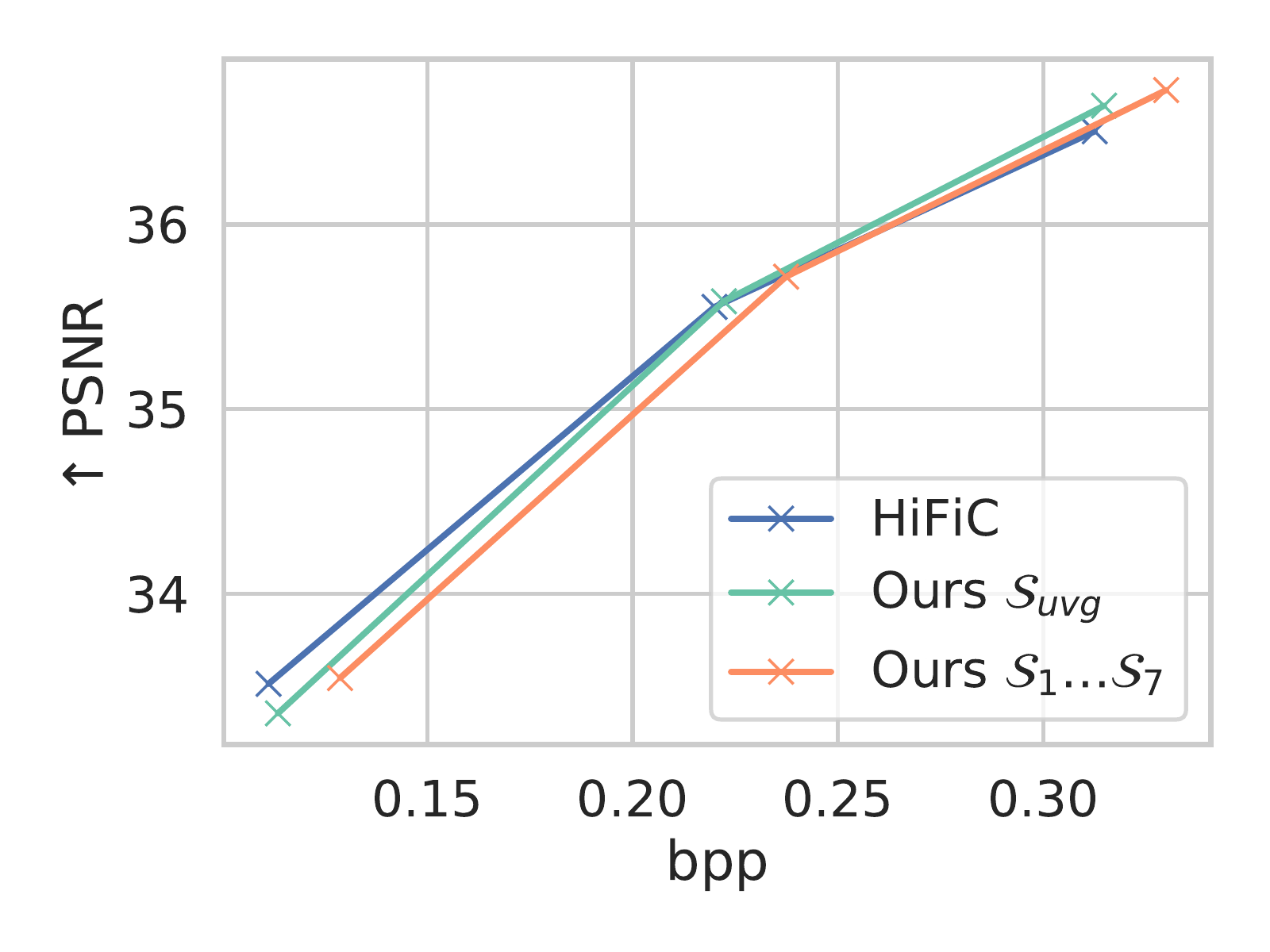}
	}
	\subfloat{
		\includegraphics[width=0.31\textwidth]{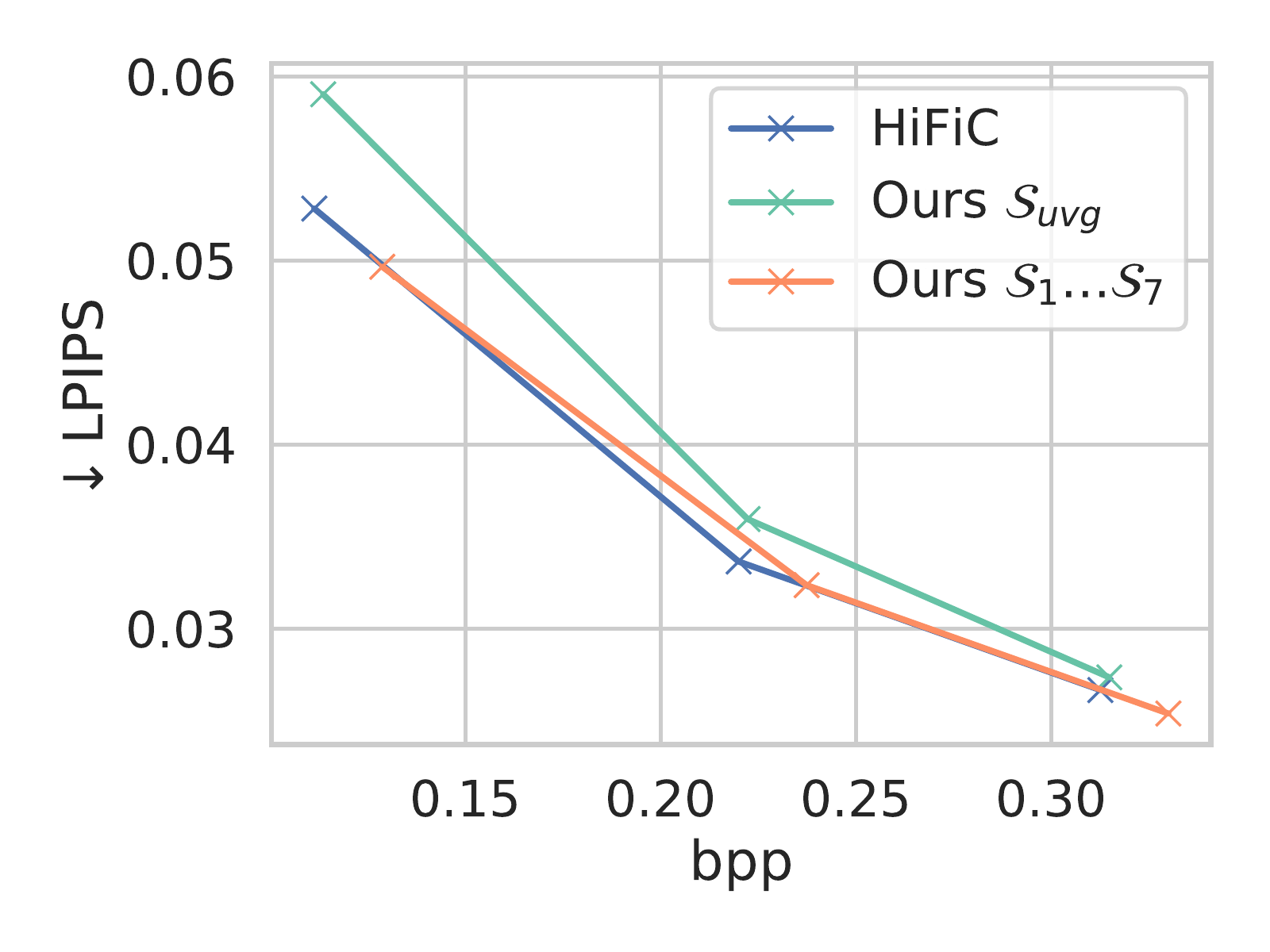}
	}
	\subfloat{
		\includegraphics[width=0.31\textwidth]{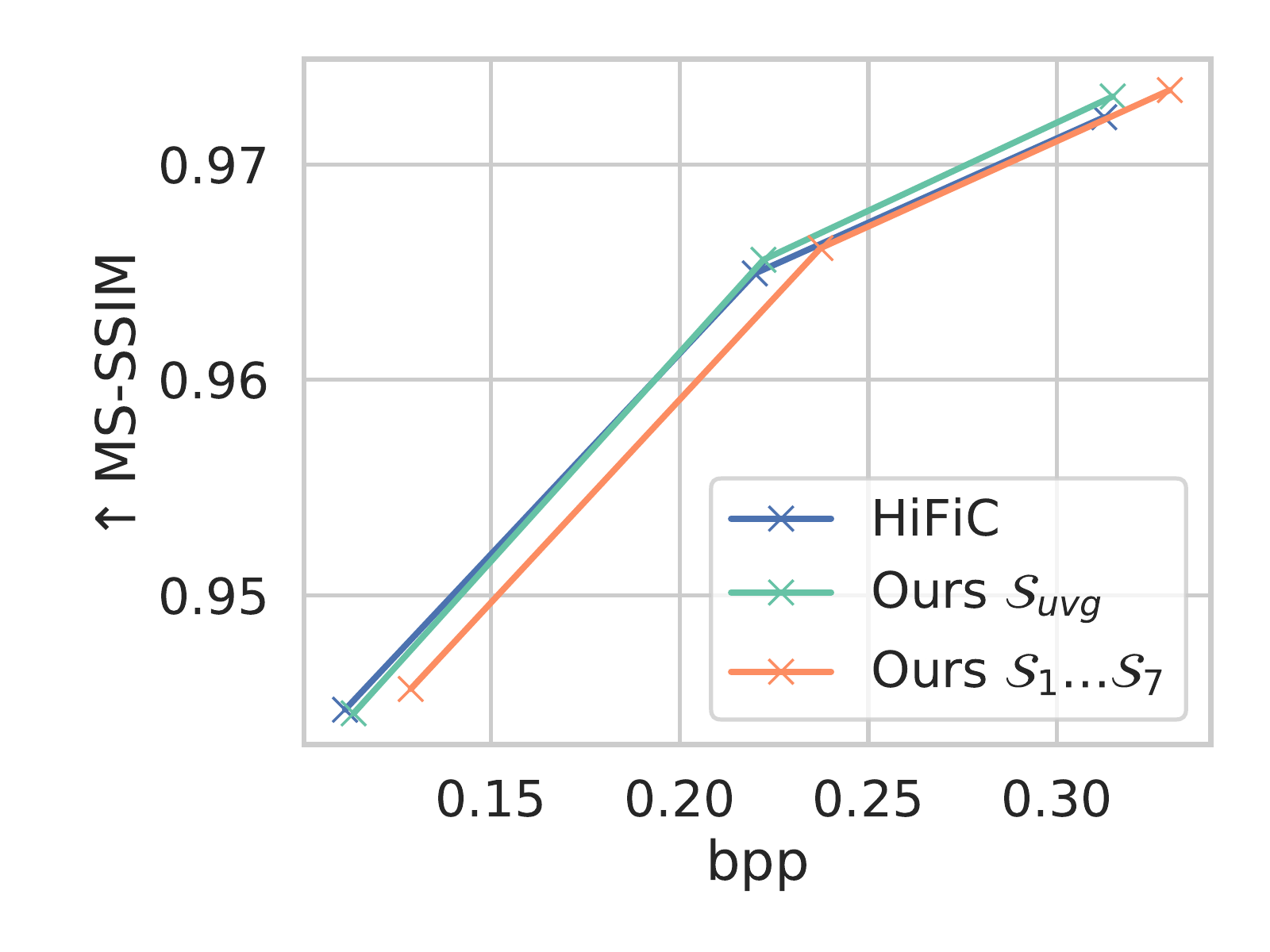}
	}
	\caption{Comparison to \emph{HiFiC}: Distilled Decoders perform similar to \emph{HiFiC} decoders. The quality of the models trained separately on each sequence is better, with the trade-off of requiring more bits for sending the weights of Micro-RN.}
	\label{fig:hific_comparison_norand}
\end{figure}

\subsection{Application: Video Compression with Latent Space Residuals}
\paragraph{Training} 
We used Adam optimizer \citep{DBLP:journals/corr/KingmaB14} with a learning
rate of $10^{-4}$ and a batch size of 4 on random crops of size $256 \times
256$.
The training of the full video compression pipeline is separated into four stages (the corresponding loss functions are defined in the Appendix):
During the first 2 phases ($100k$ steps, $50k$ steps) only the \emph{FPN} is
trained.
Phase 3 includes the entropy model for the latent residuals
for $150k$ steps.
During the last phase we optimize for multiple frames.
The back propagation of the reconstruction error through multiple frames in one
optimization step, allows the model to alleviate error accumulation
\citep{DBLP:conf/eccv/LuCZCOXG20}.
Due to memory limitations and runtime constraints, we chose $N=3$.

\paragraph{Architecture}
Our \emph{FPN}~(Figure \ref{fig:video_compression_overview})
is based on OpenDVC~\footnote{https://github.com/RenYang-home/OpenDVC}.
The flow field compression network~($E_f$, $EM_f$ and $D_f$) uses the
architecture proposed in \citep{balle2018variational} with $128$ channels.
Further, the pre-trained optical flow and motion compensation network are taken
OpenDVC.
For $E_I$ and $EM_I$ we used the pre-trained components from
$\text{\emph{HiFiC}}$~\footnote{https://hific.github.io/}. Both components are kept fix during training to leverage the knowledge of the private dataset.
For $D_I$ we either used the original \emph{HiFiC}-Decoder or our proposed distilled decoder.

%


\begin{figure}
	\centering
	\subfloat[Ground truth]{
		\includegraphics[width=0.22\textwidth]{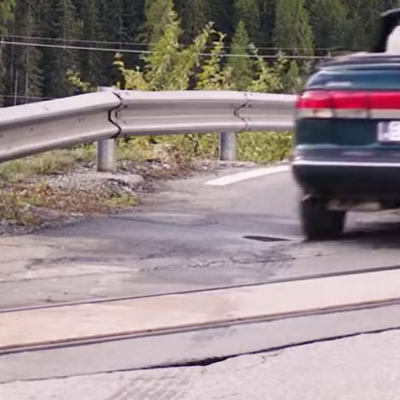}
	}
	\subfloat[LRC (Ours $\mathcal{S}_{lucid}$)]{
		\includegraphics[width=0.22\textwidth]{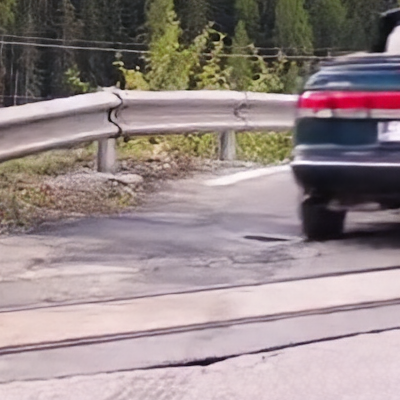}
	}
	\subfloat[LRC (Ours $\mathcal{S}_{i}$)]{
		\includegraphics[width=0.22\textwidth]{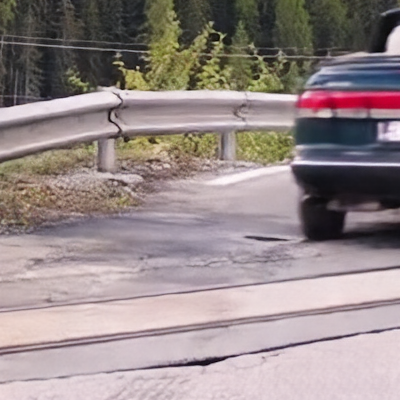}
	}
	\subfloat[OpenDVC 512]{
		\includegraphics[width=0.22\textwidth]{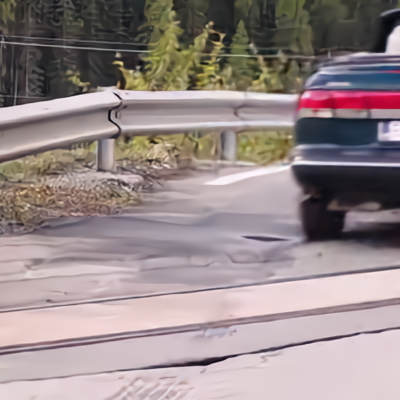}
	}
	\caption{Video Compression: Visual Comparison between LRC models (trained on $\mathcal{S}_{uvg}$ and $\mathcal{S}_{uvg}$) and  OpenDVC at the same bit rate.}
	\label{fig:comparison_alrc}
	\vspace{-1em}
\end{figure}

\begin{figure}%
	\centering
	\subfloat{
		\includegraphics[width=0.33\textwidth]{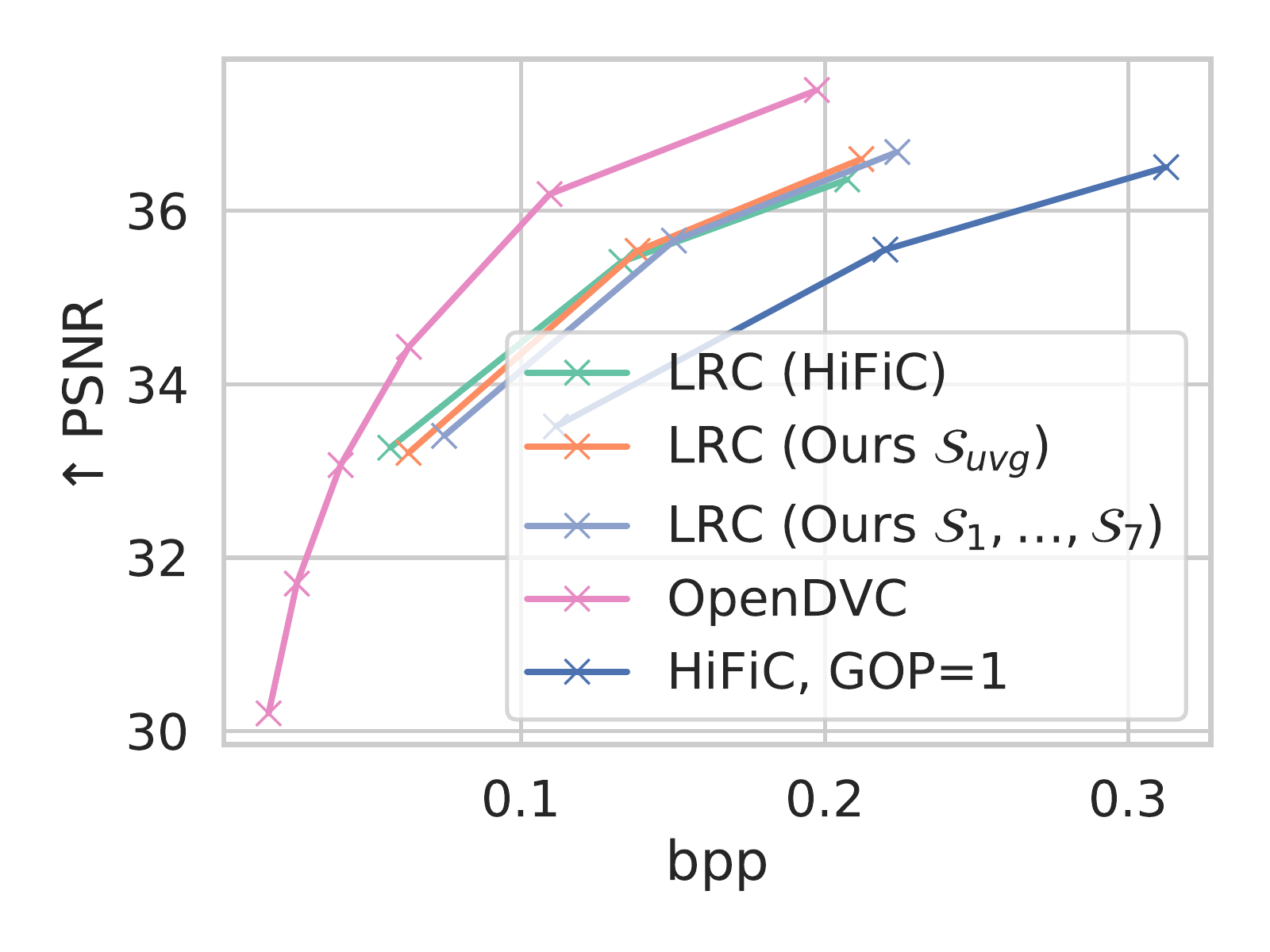}
	}
	\subfloat{
		\includegraphics[width=0.33\textwidth]{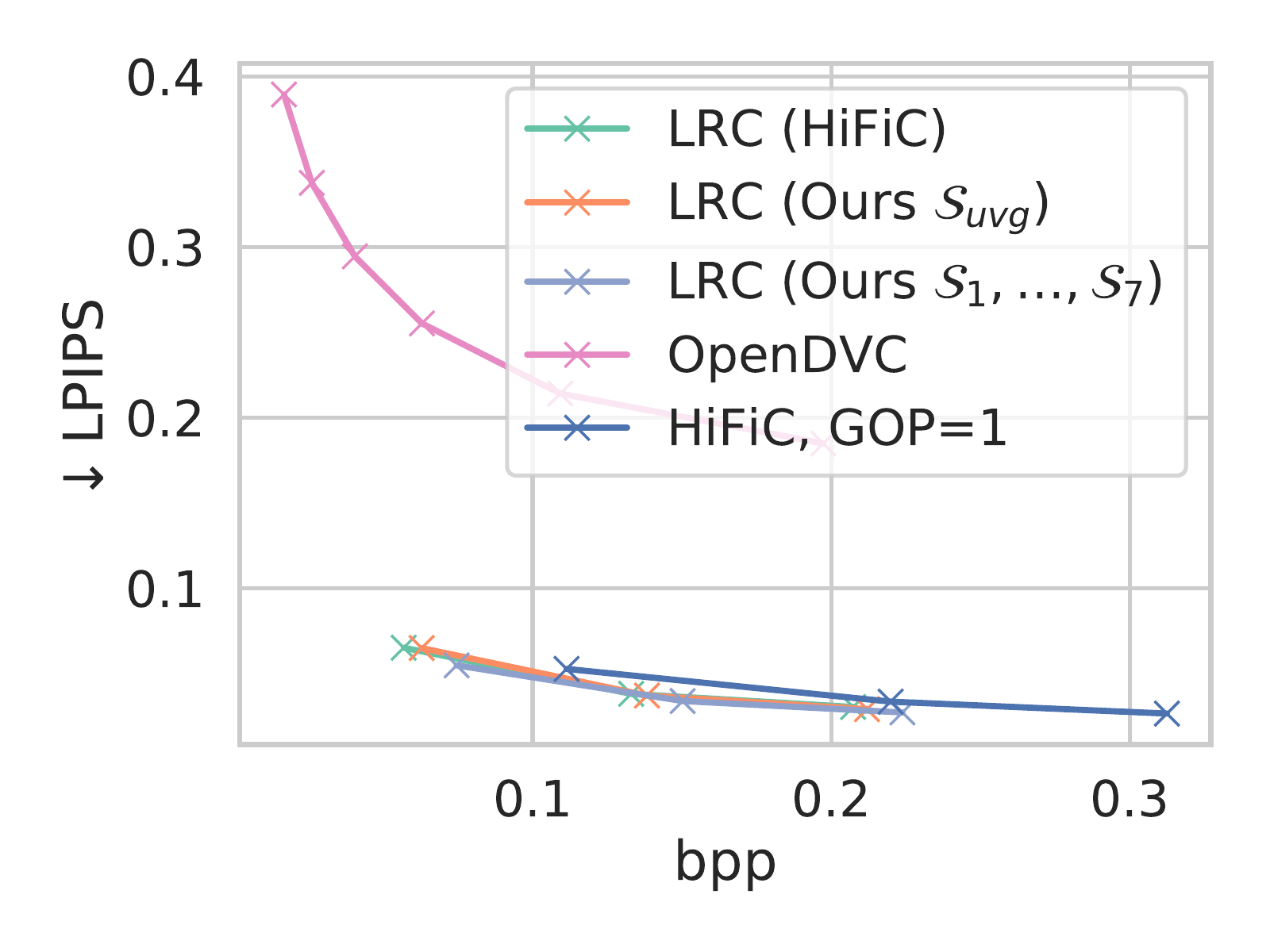}
	}
	\subfloat{
		\includegraphics[width=0.33\textwidth]{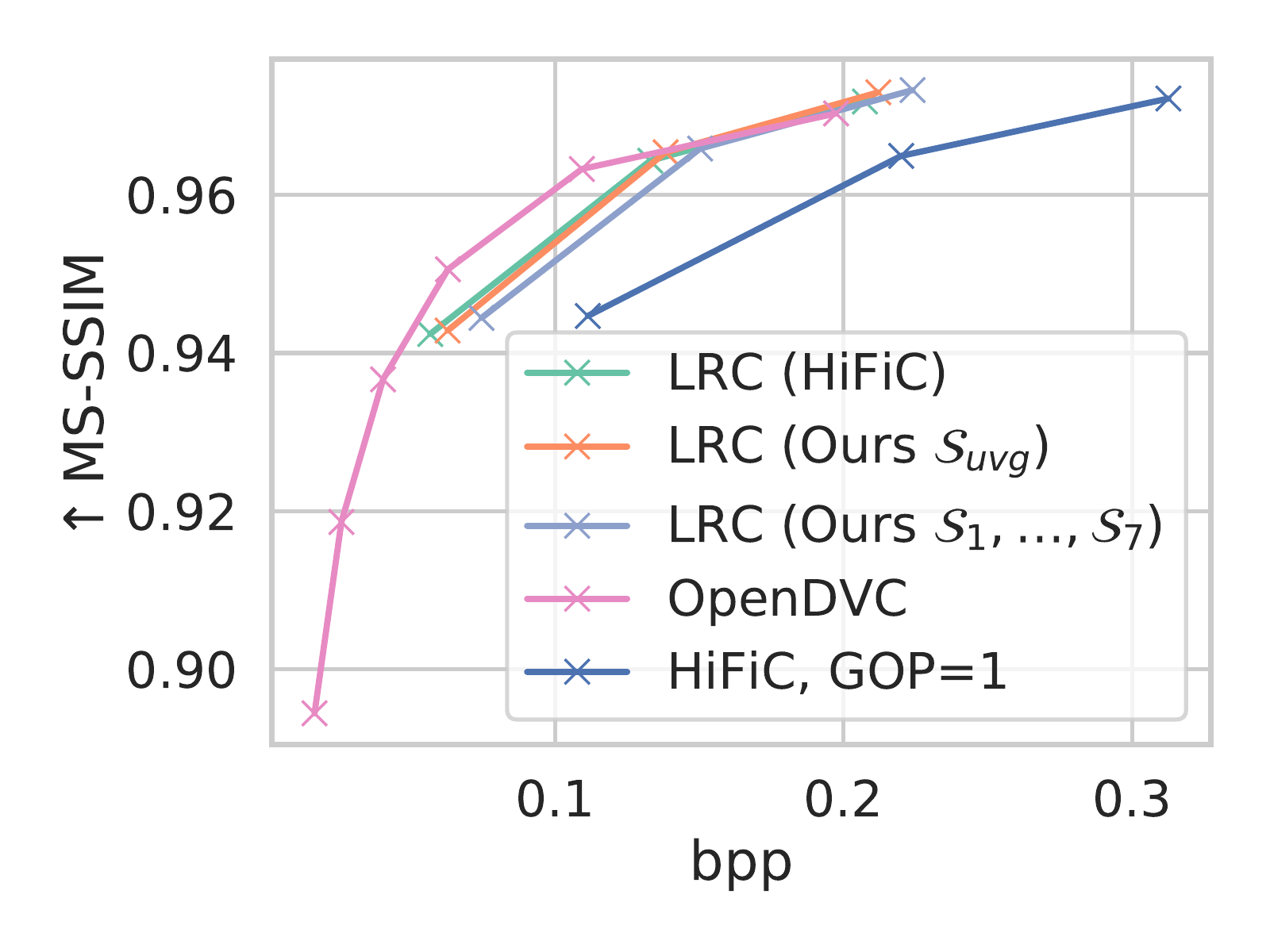}
		}
	\caption{Results for neural video codecs on UVG Dataset. 
The points on the curve correspond to (left) $\text{\emph{HiFiC}}^{Lo}$, (mid) $\text{\emph{HiFiC}}^{Mi}$ and (right) $\text{\emph{HiFiC}}^{Hi}$. If not noted, GOP=10.  
The content-specific information per subset is included in the plots and amortized over the full dataset.}
	\label{fig:hific_alrc}
\end{figure}

\vspace{-1em}
\paragraph{Results} 
We evaluated the learned models using
PSNR, LPIPS and MS-SSIM and compared the latent residual compression (LRC)
model to OpenDVC. 
Figure~\ref{fig:comparison_alrc} shows that our LRC models capture details
(e.g. textures of the street and trees), while OpenDVC oversmoothes the texture
details at the same bitrate.
To obtain the rate distortion curve~(Figure~\ref{fig:hific_alrc}) we trained a
video compression model for each of
$\text{\emph{HiFiC}}^{\{Hi, Mi, Lo\}}$.
%
It can be seen that the LRC model leverages the motion information of previous
frames. 
When compared to $\text{\emph{HiFiC}}^{Lo}$, the LRC model decreases the bitrate by almost
50\% (0.056bpp vs 0.11bpp) while keeping the reconstruction quality at a
similar level.
In case of $\text{\emph{HiFiC}}^{Hi}$, the improvement in terms of percentage is less, namely
(\textasciitilde$30$\%).
The results also show that using the distilled decoder \emph{LRC (Ours
$\mathcal{S}_{uvg}$)} produces similar quality of reconstruction and only adds
0.005bpp if encoded on the full UVG dataset~(3900 frames).
The content-specific information that has to be additionally sent is already
included in the numbers and amortized over the full dataset.
Such information is sent uncompressed. 
Also, it can be seen that if we overfit to each sequence separately the model
can better adapt to each sequence with the trade-off of sending the weights for
each subset.
This explains the increase of bpps for \emph{LRC (Ours $\mathcal{S}_1, \dots,
\mathcal{S}_7$)}.

\vspace{-0.5em}
\section{Conclusion}
\label{sec:conclusion}
In this paper, we showed how to leverage knowledge distillation in the
context of neural image and video compression. More specifically we focused
on GAN-based neural networks targeting low bitrates.
Our proposal distills the information from a big decoder and replaces it 
with a much smaller one along with sequence-specific side information.  
Altogether this allows us to have only \textasciitilde 5\% of
the original model size and to achieve 50\% reduction in decoding time.
Future work could focus on appropriate encoding of the side information
and additional methods to ensure temporal consistency.

\bibliography{KD-GAN}
\bibliographystyle{plainnat}

\newpage
\appendix
\section{Appendix}
\subsection{Knowledge Distillation for Video Compression with Latent Space Residuals}
Let consider the sequence of frames $\mathbf{x}_0, \dots, \mathbf{x}_{GOP}$.
To compress the full sequence, the I-Frame or Keyframe ($\mathbf{x}_0$) is
compressed by an image compression codec to $\hat{\mathbf{x}}_0$.
Given this compressed I-Frame, the codec computes the flow field
$\mathbf{f} = \left(\mathbf{f}_y, \mathbf{f}_x\right)$ to the next frame.
The decompressed flow field $\hat{\mathbf{f}}$ is then used to warp the
previous frame $\hat{\mathbf{x}}^{warp}_{t+1} =
\operatorname{billinear}\left(\hat{\mathbf{x}}_{t}, \hat{\mathbf{f}}\right)$.
Similar as in \citep{DBLP:conf/cvpr/LuO0ZCG19} we use a \emph{Motion
Compensation} network to fix obvious errors of the warp, which eventually
results in the final prediction $\mathbf{x}_{t+1}^{Pred}$. 
To compress the optical flow $\mathbf{f}$ we use the same auto encoder
architecture as proposed in \citep{balle2018variational}.
In our implementation we use spatial pyramid network
\citep{DBLP:journals/corr/RanjanB16} for flow field estimation.


To compute the latent residuals, we first encode both frames, $\mathbf{x}_{t+1}^{Pred}$ and $\mathbf{x}_{t+1}$ with the pre-trained \emph{HiFiC} encoder and take the difference of the encodings $\mathbf{r}_{t+1} = \mathbf{y}_{t+1} - \mathbf{y}^{Pred}_{t+1}$. The probability distribution of the residuals is then learned with a Scale-Hyperprior \citep{balle2018variational}. By adding the decompressed residuals and the encodings of the prediction $\hat{\mathbf{y}}_{t+1} = \hat{\mathbf{r}}_{t+1} + \mathbf{y}^{Pred}_{t+1}$ we obtain the latents of the next frame $\hat{\mathbf{x}}_{t+1}$.\\


The training of the model is separated into four stages. In the first stage, we only train the motion vector compression network. The loss we optimize is defined as:

\begin{equation}
	\mathcal{L}_{warp} = \lambda_{w} r\left(\hat{\mathbf{w}}_{t+1}\right) + \operatorname{MSE}\left(\mathbf{x}_{t+1}, \hat{\mathbf{x}}^{warp}_{t+1}\right)
	\label{eq:loss_phase1}
\end{equation}

where $\mathbf{w}_{t+1}$ is the encoding of flow field $\mathbf{f}$ and $\lambda_{w}$ is a hyperparameter controlling the trade-off between the distortion term and the rate term $r\left(\hat{\mathbf{w}}_{t+1}\right)$.

In the second phase, the \emph{Motion Compensation} network is trained by optimizing the following loss:

\begin{equation}
	\mathcal{L}_{mc} = \lambda_{w} r\left(\hat{\mathbf{w}}_{t+1}\right) + k_M \operatorname{MSE}\left(\mathbf{x}_{t+1}, \mathbf{x}_{t+1}^{Pred}\right) + k_M  L_1\left(\mathbf{y}_{t+1}, \mathbf{y}^{Pred}_{t+1}\right) .
	\label{eq:loss_phase2}
\end{equation}

Note, that the $L_1$ forces the model to learn predictions with latents close to latents of the ground truth image $\mathbf{x}_{t+1}$.

The loss of the third phase is defined as:
\begin{equation}
\mathcal{L}_{step} = \lambda_{w} \left(r\left(\hat{\mathbf{w}}_{t+1}\right) + r\left(\hat{\mathbf{r}}_{t+1}\right)\right) + k_M\operatorname{MSE}\left(\tilde{\mathbf{x}}_{t+1}, \hat{\mathbf{x}}_{t+1}\right) + k_p d_p\left(\mathbf{x}_{t+1}, \hat{\mathbf{x}}_{t+1}\right) .
\label{eq:loss_phase3}
\end{equation}

Important to note, in this phase we make use of the teacher-decoder.
We minimize the loss between the compressed P-Frame $\hat{\mathbf{x}}_{t+1}$ and the \emph{HiFiC} compressed frame $\tilde{\mathbf{x}}_{t+1}$. 

In the fourth and final phase we optimize for $N = 3$ frames in one optimization step, to consider more temporal information and alleviate error accumulation \cite{DBLP:conf/eccv/LuCZCOXG20}:

\begin{equation}
\mathcal{L}_{final} = \lambda \sum^N_{i=1} \left(r\left(\hat{\mathbf{w}}_{i}\right) + r\left(\hat{\mathbf{r}}_{i}\right)\right) + \sum^N_{i=1} k_M \operatorname{MSE}\left(\tilde{\mathbf{x}}_{i}, \hat{\mathbf{x}}_{i}\right) + k_p d_p\left(\mathbf{x}_i, \hat{\mathbf{x}}_i\right),
\label{eq:loss_phase4}
\end{equation}

\end{document}